\documentclass[reprint]{revtex4-2}
\usepackage{graphicx}
\usepackage{amsmath}
\usepackage{graphicx}
\usepackage{makecell}
\usepackage{multirow}
\usepackage[english]{babel}
\usepackage{subfigure} 

\begin{document}

\title{Modeling Torque Induced Alignment in a Dusty Plasma System}

\author{Benny Rodriguez Saenz}
\author{ Diana Jimenez Marti}
\author{Lorin Swint Matthews}
\author{Truell W. Hyde}

\affiliation{Center for Astrophysics, Space Physics, and Engineering Research,
    Baylor University, One Bear Place 97316, Waco, Texas 76798-7316, USA}

\begin{abstract}
    Irregular dust aggregates immersed in plasma sheaths experience several
    orientation-dependent torques that can modify their rotational dynamics and
    stability. Here, we investigate the rotational dynamics of charged irregular
    aggregates under conditions representative of a GEC rf plasma cell using
    self-consistent numerical simulations. The aggregates rotate freely in a
    unidirectional sheath electric field that drives an ion flow, allowing the
    torque contributions acting on the aggregate to be evaluated throughout the
    motion. The results show that the sheath electric field is the main driver
    of rotation and aligns the aggregate electric dipole moment with the sheath
    field direction. The ion wake modifies this alignment: its axial field
    component produces an opposing torque, while its transverse components
    introduce a destabilizing contribution that leads to small oscillations
    about the equilibrium orientation. The rotational equilibrium is described
    by an interaction energy well whose spring constant and depth increase with
    the sheath electric field magnitude, indicating stronger alignment and
    greater resilience to angular perturbations at higher fields. A second order
    multipole expansion of the aggregate ion interaction shows that the dipolar
    term governs the ion contribution to the aligning torque, supporting a
    dipole ion approximation across the examined conditions. These results
    identify the sheath electric field as the principal stabilizing mechanism
    for irregular aggregate rotation and clarify how ion wake fields perturb the
    equilibrium orientation.
\end{abstract}

\maketitle

\section{Introduction}

Dusty plasmas consist of charged solid particles embedded in an ionized gas
environment. The dust grains become charged, and their presence modifies the
local plasma environment, with effects that depend on particle geometry
\cite{merlinoDustyPlasmasLaboratory2004}. Charged dust grains interacting with
electromagnetic fields are of major interest in astrophysical and laboratory
plasma environments. In the interstellar medium and protoplanetary disks,
aligned non-spherical dust grains produce polarized thermal emission that traces
magnetic field geometry and strength, with recent high-resolution disk
observations even enabling three-dimensional magnetic field reconstruction
\cite{anderssonInterstellarDustGrain2015,ohashiObservationallyDerivedMagnetic2025}.
In the semiconductor industry, dust contamination is responsible for the failure
of microchips created by plasma etching \cite{selwynLaserDiagnosticStudies1989},
whereas erosion of plasma-facing components in fusion tokamaks produces dust
that can promote disruption events, with implications for both operational
safety and device efficiency \cite{ratynskaiaDustPowderFusion2022,
    rubelDustGenerationTokamaks2018}. Plasma environments can also influence the
formation and growth of dust particles. In low-pressure processing plasmas,
nanoparticles can nucleate and grow through coagulation and molecular sticking,
with the growth dynamics strongly affected by particle charging and plasma
conditions \cite{kortshagenGenerationGrowthNanoparticles1999,
    kortshagenModelingParticulateCoagulation1999}. In cryogenic laboratory plasmas,
elongated and fractal-like water-ice grains have been observed to form
spontaneously, levitate in the sheath, and align parallel to the sheath electric
field \cite{chaiFormationAlignmentElongated2015}.

There are several physical phenomena that can produce aligning torques for
non-spherical grains such as electric fields in the plasma sheath, ion wakes in
flowing plasma, mechanical collisions with gas particles or plasma particles and
magnetic fields. Experiments on irregular cylindrical particles levitated in a
plasma sheath have demonstrated that such particles exhibited preferred
orientations and a stable rotational equilibrium
\cite{annaratoneLevitationCylindricalParticles2001,
    ivlevRodlikeParticlesGas2003}. Depending on grain size and the location within
the sheath, both vertical and horizontal alignments have been observed. These
orientations are driven by sheath electric field gradients, which create a
torque by unbalancing the electrostatic lifting force along the rod relative to
its weight. Ion drag and ion wakes can further influence this alignment. Ion
wakes arise when an electric field drives a steady ion flow relative to negative
dust grains. The streaming ions interact with this negative charge, producing an
enhanced region of positive charge downstream of a dust grain. This anisotropic
ion distribution within the plasma can significantly influence the dynamics of
other dust particles. Even though particle charge plays a dominant role in dusty
plasmas, other mechanical interactions must be considered. Numerical simulations
of astrophysical dusty plasmas in which paramagnetic irregular aggregates are
immersed in a gas flow found that shape-dependent and material-dependent torques
can orient the grains toward stable alignment configurations
\cite{reisslMechanicalAlignmentDust2023}. In these systems, the resulting
alignment can arise from purely mechanical interactions with the gas flow, but
may also involve a competition with magnetic torques associated with the
magnetic properties of the dust grains.

To isolate the role of plasma flow and geometry, the interaction of fixed,
elongated rod-like particles was investigated numerically
\cite{milochInteractionTwoElongated2009}. In this work, the simulations
demonstrated that the resulting ion wakes and interaction potentials are highly
non-linear and depend strongly on the rods' relative distances and orientations.
For rods aligned perpendicularly to the flow at distances comparable to the
electron Debye length, a unified wake forms, channeling and enhancing the ion
density between the grains through an electrostatic lensing effect. Furthermore,
varying the rods' inclination angles relative to the flow produces asymmetric
ion fluxes and potential distributions, which consequently exert rotational
torques on the elongated grains.

The non-uniform shape of an aggregate produces a non-uniform charge distribution
on its surface. The outermost regions collect a larger fraction of the charge,
whereas less charge is found near the core of the aggregates
\cite{matthewsDiscreteStochasticCharging2018, matthewsChargingFractalDust2007}.
A multipole expansion approach can be used to describe this charge distribution;
however, to determine the appropriate number of multipole moments to include in
the expansion, the plasma conditions and distance from the aggregate must be
considered \cite{matthewsMultipoleExpansionsAggregate2016}. Such anisotropic
charge distribution can influence the forces and torques acting on an aggregate.

Modeling the formation of ion wakes and their interaction with irregular grains
requires resolving the plasma response self-consistently, since the ion density
distribution around the grain depends on multiple factors such as the plasma
flow, the particle geometry, and its charge. It has been shown that in flowing
plasmas, the anisotropic potential of prolate spheroids deflects streaming ions
asymmetrically, generating transverse ion drag components
\cite{krasheninnikovDragForceNonspherical2011}. Even in stationary plasmas,
longitudinally asymmetric grains (e.g., drop-like shapes) can experience a net
force because their asymmetric shape breaks the symmetry of ion scattering
\cite{krasheninnikovForceExertedNonspherical2024}. The interaction of charged
irregular dust grains with the surrounding plasma also leads to asymmetric
mechanical effects, producing translational and rotational motions that are
challenging to predict \cite{asnazChargingIrregularlyShaped2018}.

Experiments have been conducted to produce aggregates grains and measure their
net charge and dipole moment, the lack of control over the aggregate shapes
constitutes a limitation in reproducing specific effects.  In most dusty plasma
laboratories, data are collected by high speed cameras in a two-dimensional
format. Reconstructing an aggregate's three-dimensional shape requires some
simplifying assumptions, such as approximating the aggregate's shape as a
collection of equal sized spheres (monomers) and restricting the motion of
aggregates to the plane of view when estimating their dipole moments
\cite{yousefiMeasurementNetElectric2014}.

Despite these shape-dependent effects, most numerical models, such as those used
to simulate dust grain transport in fusion devices, still assume the particles
are spherical \cite{bacharisDustTokamaksOverview2010,
    delzannoSurvivabilityDustTokamaks2014, liuSimulationDustGrain2017,
    martinModellingDustTransport2008, pigarovDustparticleTransportTokamak2005}.
Since the orientation of an aggregate modifies its collisional cross-section as
it moves through the plasma, accounting for irregular grain shapes is essential
to accurately predict the impact of dust contaminants in these regimes. In
particular, it is unclear whether irregular aggregates align with the sheath
electric field through their electric dipole moment, or instead reach a
different equilibrium orientation due to ion wake-driven torque and other
effects. Therefore, quantifying the torque balance and rotational equilibrium of
irregular dust grains motivates a fundamental investigation under
well-controlled plasma conditions.

In this work, we model the charging process and rotational dynamics of irregular
aggregates in the presence of a unidirectional electric field driving a directed
ion flow. The simulations are performed using the DRIAD code
\cite{matthewsDustChargingDynamic2020}, which self-consistently models the ion
dynamics and charging of irregular dust grains. Aggregates are allowed to rotate
freely, enabling the analysis of torque balance and equilibrium orientations.
Rotational equilibrium configurations are analyzed based on a multipole
expansion of the interaction energy between the aggregate and the surrounding
ions, which allows dominant contributions to the rotational dynamics to be
identified. As a representative experimental configuration, conditions
corresponding to a GEC rf cell are considered
\cite{yousefiMeasurementNetElectric2014}. However, the physical mechanisms
arising from the interaction of irregular grains with streaming ions are general
and relevant to other plasma environments, such as the edge of fusion devices or
astrophysical systems, where directed ion flows interact with dust particles.

This manuscript is structured as follows. Section \ref{sec:Methods} details how
we model the aggregates, the aggregate charging, the forces acting on the
aggregates causing them to rotate, and the interaction energy of the aggregate.
In Section \ref{sec:Results} we present a detailed analysis of the results for a
single aggregate, showing that the aggregate's electric dipole aligns with the
sheath electric field and that the interaction with the ion electric field is
the dominant factor in altering this alignment. In Section IV, we present
results for aggregates with different shapes, from very elongated to almost
spherical.

\section{Methods}

\label{sec:Methods}

\subsection{Modeling aggregates}

In this study, aggregates are modeled as rigid bodies composed of equal-radius
monomers, created using the aggregate-builder code
\cite{matthewsFormationCosmicDust2007}. Each aggregate is characterized by the
elongation factor:

\begin{equation}
    \xi = \sqrt{ \frac{1}{3} \left( \frac{1}{I_1/I_2} + \frac{1}{I_1/I_3} +
        \frac{1}{I_2/I_3} \right) - 3 },
    \label{eq:elongation factor}
\end{equation}
where  $I_1, I_2, \text{and} \,I_3$ are the principal moments of inertia in
ascending order. The elongation parameter $\xi$ quantifies how different an
aggregate is from an isotropic body, where $\xi = 0$ corresponds to a perfectly
isotropic object. The set of aggregates was selected to range from highly
elongated geometries ($\xi=2.02$) to nearly spherical shapes ($\xi=0.1$). In
this context, $\xi$ is only used to indicate relative morphological diversity
among the aggregates considered.

\begin{figure}
    \centering
    \includegraphics[width=\columnwidth]{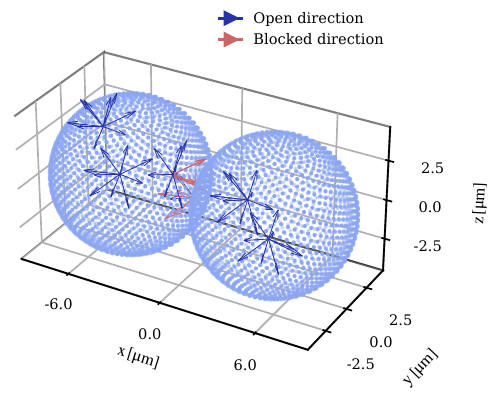}
    \caption{Monomer surface points used to calculate LOS factor for electron
        current. Arrows show incident-direction sampling, retaining only outward
        directions. Purple and orange arrows indicate open and blocked
        directions, respectively.}
    \label{fig:monomer}
\end{figure}

\subsection{Aggregates charging process}

As the aggregates are immersed in a plasma environment, they naturally acquire
electric charge on their surfaces through the collection of plasma species
(electrons and ions). The resulting aggregate charge distribution interacts with
both the sheath electric field and the ion electric field.

An aggregate's charging and dynamical evolution are modeled using DRIAD
\cite{matthewsDustChargingDynamic2020}, a molecular dynamic code which solves
the equation of motion of both ions and aggregates on their respective time
scales. The charge of the aggregate is calculated self-consistently depending on
plasma conditions. To account for the different time scales involved, aggregates
are held fixed while ion dynamics are advanced $N(\Delta t_i)$ ion time steps.
Subsequently, the ions are frozen, and the aggregate's position (orientation) is
updated. Ions are modeled as super-ions to reduce computational cost. Each
super-ion represents a cluster of $n_i\,V_{\text{sim}}/N_i$ physical ions with
the same charge-to-mass ratio as a single physical ion, where $n_i$ is the ion
number density, $V_{\text{sim}}$ is the simulation volume and $N_i$ is the
number of super-ions used in the simulation. Super-ions collected by the
aggregate or that leave the simulation volume are reinjected on the simulation
boundary according to the scheme in \cite{hutchinsonIonCollectionSphere2002}.

In the model implemented in DRIAD, ion motion is integrated to the aggregate
surface, therefore the number of ions collected by each monomer can be directly
quantified. Electrons, on the other hand, are assumed to be Boltzmann
distributed and approach the aggregate surface isotropically. We model the
electron current to each monomer in the aggregate by the OML current, which is
modified to take into account the aggregate geometry:

\begin{equation}
    \begin{aligned}
         & I^m_e =
        I_0 \exp{\left(-\frac{q_e V_m}{k_B T_e}\right)} \text{LOS}_m ,                                        \\
         & I_0 = n_e q_e A_p \frac{\Delta \Omega}{\pi} \left( \frac{k_B T_e}{2 \pi m_e}\right)^{\frac{1}{2}}.
    \end{aligned}
    \label{eq:electron current}
\end{equation}
Here $n_e$ is the plasma electron density, $q_e$  is the electron charge, $k_B$
is the Boltzmann constant, $T_e$ is the electron temperature, $m_e$ is the
electron mass, and $V_m$ is the electric potential on the surface of monomer
$m$. The factor $LOS_m$ accounts for the fact that electrons can only approach a
monomer's surface from directions that are not blocked by other monomers in the
aggregate, or open Lines of Sight (LOS) as described in
\cite{matthewsDiscreteStochasticCharging2018}. To model this effect, each
monomer $m$ is discretized into $N_p$ points distributed quasi-uniformly on the
surface of a sphere of radius $R_m$ using the Fibonacci spiral algorithm as
illustrated in Fig.~\ref{fig:monomer}. This algorithm ensures an efficient and
homogeneous coverage of the sphere, minimizing clustering at the poles
\cite{gonzalezMeasurementAreasSphere2010}. Each point defines a patch with area
$A_p=4\pi R_m^2/N_p$ \cite{matthewsDiscreteStochasticCharging2018}. To determine
the open incident directions, test directions $\hat{t}$ emanating from the
center of each patch (the Lines of Sight) are determined to be blocked if they
intersect any other monomer in the aggregate ($\text{LOS}^{p}_{m}(t)=0$) or open
if they are not blocked ($\text{LOS}^{p}_{m}(t)=1$), as shown in
Fig.~\ref{fig:monomer}. The LOS factor for each monomer $m$ is then given by:

\begin{equation}
    \label{eq:LOS factor}
    \text{LOS}_m=\sum^{N_p}_p \sum^{N_t}_t
    \text{LOS}^{p}_{m}(t)\cos^{p}_{m}(\theta_t) \text{,}
\end{equation}

\noindent where $\theta_t$ is the angle between the surface normal of the patch
and the test direction $\hat{t}$. $\Delta\Omega = 4\pi / N_t$ is the solid angle
associated with each test direction. $\text{LOS}_m$ is calculated once at the
beginning of the simulation with $N_p=200$ and $N_t=10 000$ to ensure sufficient
accuracy.

The charge variation on monomer $m$ due to electron collection during an ion
time step ($\Delta t_i = 10^{-8}\,\text{s}$) is given by $\Delta Q^e_m =
    I^e_m\Delta t_i$. As ions are tracked individually in the super-ion model, the
charge increment $\Delta Q^i_m$ is computed as the amount of charge collected by
the surface due to super-ion collisions during one ion time step. Thus, the
total charge variation $\Delta Q_m$ on monomer $m$ is:
\begin{equation}
    \Delta Q_m = \Delta Q^e_m + \Delta Q^i_m
\end{equation}

Because a single super-ion represents approximately $\sim 100$ physical ions,
the collection of an individual super-ion produces large charge fluctuations on
each monomer. To reduce these fluctuations, the charge collected is smoothed  on
each aggregate time step ($\Delta t_n=10^{-5}\, \text{s}$) as implemented in
\cite{matthewsDustChargingDynamic2020}:

\begin{equation}
    Q_{m}(t_n) = 0.95 \, Q_{m}(t_{n-1}) + 0.05 \, Q^{\text{avg}}_m(t_n),
\end{equation}
here, $t_n$ and $t_{n-1}$ denote the current and previous aggregate time step
and $Q^{\text{avg}}_m(t_n)$ is the average charge of the monomer during the
$N(\Delta t_i)$ ion time steps. The total charge on the aggregate is
$Q=\sum_mQ_m$ and the aggregate dipole moment is:

\begin{equation}
    \label{eq:dipole moment}
    \vec{p}=\sum_m Q_m\vec{r}_m,
\end{equation}
where $\vec{r}_m$ is the position of the monomer center with respect to the
aggregate center of mass.

\subsection{Rotational dynamics}
\subsubsection{Reference frames}

\begin{figure}
    \centering
    \includegraphics[scale=1]{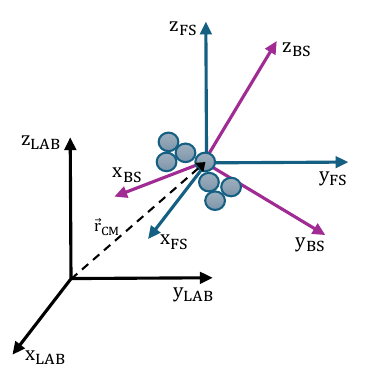}
    \caption{Definition of reference frames. All rotations and torques are fully
    three-dimensional. Black: Laboratory system (LAB). Blue: Fixed system (FS).
    Purple: Body system (BS). The aggregate's center of mass is at
    $\vec{r}_{\text{CM}}$ in the LAB frame.}
    \label{fig:reference frames}
\end{figure}

The aggregates experience various forces from the plasma environment as well as
torques due to their non-uniform charge and mass distribution. To accurately
describe the rotational motion, three reference frames are defined as shown in
Fig.~\ref{fig:reference frames}. In the simulation, the center of mass of each
aggregate is fixed in the LAB system, and the aggregate is allowed to rotate
about this point.

\begin{itemize}
    \item \textbf{Laboratory System (LAB)}: A Cartesian coordinate system whose
          origin coincides with the center of the simulation volume. All
          simulation parameters are defined in this frame. This system is
          introduced to distinguish between the origin of the simulation volume
          and the center of mass of the aggregate. These two origins in general
          do not coincide.

    \item \textbf{Fixed System (FS)}: A coordinate system parallel to the LAB
          system with its origin at the aggregate center of mass. Rotational
          motion is defined with respect to this origin.

    \item \textbf{Body System (BS)}:  A coordinate frame defined by the
          aggregate's principal axes with its origin at the aggregate center of
          mass. This frame rotates with the aggregate and the Euler equations of
          motion are integrated in this body-fixed frame.
\end{itemize}

\subsubsection{Acting forces}

To quantify the rotational effects of the different forces acting on the
aggregates, their contributions to the total force acting on an individual
monomer must be computed.

The forces acting on each monomer $m$ are:

\begin{itemize}
    \item \textbf{Sheath electric force} ($\vec{F}^m_e$): Arises from the
          interaction between the charge on monomer $m$ and the electric field
          in the plasma sheath $\vec{E}_{\text{sheath}}$:

          \begin{equation}
              \vec{F}^m_e = Q_m \vec{E}_{\text{sheath}}.
              \label{eq:electric force}
          \end{equation}

    \item \textbf{Ion drag (orbital) force} ($\vec{F}^m_{i,\text{orb}}$):
          Corresponds to the electrostatic interaction between the ions and
          monomer $m$ with charge $Q_m$, averaged over one aggregate time step:

          \begin{equation}
              \begin{split}
                  \vec{F}^m_{i,\text{orb}} = & \frac{ Q_m q_i}{4\pi\epsilon_0 N(\Delta t_i)} \sum^{N(\Delta t_i)}_{j=1} \sum^{N_i}_{i=1}
                  \frac{1 + |\vec{r}_m - \vec{r}_i| / \lambda_{D_e}}{|\vec{r}_m - \vec{r}_i|^3}                                          \\
                                             & \exp{ \left (-|\vec{r}_m - \vec{r}_i| / \lambda_{D_e} \right )} (\vec{r}_m - \vec{r}_i).
              \end{split}
              \label{eq:orbital force}
          \end{equation}

          Here, $\epsilon_0$ is the vacuum permittivity, $q_i$ is the superion
          charge, $N(\Delta t_i)$ is the number of ion updates per aggregate
          time step, $\vec{r}_m$ and $\vec{r}_i$ are the positions of monomer
          $m$ and superion $i$ in the LAB frame, $\lambda_{D_e}$ is the electron
          Debye length and $N_i$ is the number of super-ions in the simulation
          \cite{matthewsDustChargingDynamic2020}.

    \item \textbf{Ion drag (collisional) force} ($\vec{F}^m_{i,\text{coll}}$):
          Represents the average force exerted by direct mechanical collisions
          of superions with monomer $m$ during one aggregate time step:

          \begin{equation}
              \vec{F}^m_{i,\text{coll}} = \frac{m_i}{N(\Delta t_i)\Delta t_{i}}
              \sum^{N_{i,\text{coll}}}_{i} \vec{v}_m^{\,i,\text{coll}}.
              \label{eq:coll force}
          \end{equation}

          Here, $m_i$ is the superion mass and $\Delta t_{i}$ is the ion time
          step. The vector $\vec{v}_m^{\,i,\text{coll}}$ corresponds to the
          velocity of an ion at the time of collision with monomer $m$. The sum
          is over the superions that collide with monomer $m$
          \cite{matthewsDustChargingDynamic2020}.

    \item \textbf{Frictional force} ($\vec{F}^m_f$): This is the drag exerted by
          neutral gas which opposes the motion of the aggregate:

          \begin{equation}
              \begin{aligned}
                   & \vec{F}^m_f = S_m \beta_m M_m \vec{v}_m ,                                                     \\
                   & S_m = \frac{1}{{N_p} N_{\text{t}}} \sum^{N_p}_p \sum^{N_{\text{t}}}_t \text{LOS}^{p}_{m}(t) , \\
                   & \beta_m = \delta \frac{4}{3} \frac{R^2_m P_n}{M_m} \sqrt{\frac{8 \pi m_n}{k_B T_n}} .
              \end{aligned}
              \label{eq:frictional force}
          \end{equation}

          The dimensionless parameter $S_m \in [0, 1]$ is the shielding factor,
          which accounts for the reduced exposed area of the monomer due to
          obstruction by adjacent particles within the aggregate. The quantities
          $M_m$, $R_m$, and $\vec{v}_m$ are the mass, radius, and velocity of
          monomer $m$.  $\beta_m$ is the drag coefficient, where $P_n$ and $T_n$
          correspond to the neutral gas pressure and temperature, $m_n$ is the
          mass of a neutral gas particle, and $\delta$ is a material-dependent
          coefficient, $\delta \in [1, 1.44]$, with experimental evidence for
          melamine formaldehyde in argon supporting $\delta = 1.44$
          \cite{melzerPhysicsDustyPlasmas2019}.

    \item \textbf{Brownian force} ($\vec{F}^m_B$): Models random kicks due to
          the thermal motion of neutral gas particles
          \cite{matthewsDustChargingDynamic2020}:

          \begin{equation}
              \begin{aligned}
                   & \vec{F}^m_B = M_m \zeta \sum_{i=1}^{3} \chi^i \hat{e}_i       ,          \\
                   & \zeta = \sqrt{\frac{2 S_m \beta_m k_B T_n}{M_m \Delta t_{\text{agg}}}} .
              \end{aligned}
              \label{eq:brownian force}
          \end{equation}
          Here, $\chi^i$ is a normally distributed random number between $-3$
          and $3$ and $\hat{e}_i$ are the Cartesian unit vectors in the $i$-th
          direction.

\end{itemize}

It should be noted that the gravitational force is excluded in our model, as its
magnitude is several orders of magnitude smaller than all other forces acting on
the aggregate.

\subsubsection{Rigid body equations of motion}

To simplify the analysis we fix the center of mass and only the rotational
degrees of freedom are allowed to evolve. The torque acting on monomer $m$ with
respect to the aggregate center of mass can be written as

\begin{equation}
    \vec{\tau}_m = \vec{r}_m \times \vec{F}_m.
\end{equation}
The quantities $\vec{r}_m$ and $\vec{F}_m$ represent the position of monomer $m$
and the total force acting on it in the FS, respectively. Therefore, the total
torque on the aggregate  is written as $\vec{\tau}^{\text{FS}} = \sum_m
    \vec{\tau}_m$, where the sum is over all monomers.

To determine the time evolution of the aggregate's orientation, rotational
dynamics are solved using the Euler equations of motion
\cite{greinerClassicalMechanicsSystems2003}. These equations are naturally
written in the BS, where the inertia tensor is diagonal. The torques are
transformed from the FS to the BS using the time-dependent rotation matrix
$\overleftrightarrow{R}$, defined by the eigenvectors of the inertia tensor
$\overleftrightarrow{I}$ \cite{thorntonClassicalDynamicsParticles2008}:

\begin{equation}
    \vec{\tau}^{\text{BS}}_j (t_n) = \overleftrightarrow{R}(t_n) \,
    \vec{\tau}^{\text{FS}}_j (t_n)
    \label{eq:body system torques}
\end{equation}

The rotational dynamics in the BS are given by Euler's equations
\cite{greinerClassicalMechanicsSystems2003}:

\begin{equation}
    \begin{aligned}
        \dot{\omega}_x (t) = [\tau_x (t) +
        (I_{yy} - I_{zz}) \omega_y (t) \omega_z (t)]/I_{xx},
        \\
        \dot{\omega}_y (t) = [\tau_y (t) +
        (I_{zz} - I_{xx}) \omega_z (t) \omega_x (t)]/I_{yy},
        \\
        \dot{\omega}_z (t) = [\tau_z (t) +
        (I_{xx} - I_{yy}) \omega_x (t) \omega_y (t)]/I_{zz},
    \end{aligned}
    \label{eq:euler equations}
\end{equation}
where ${\dot{\omega}}_j$ and $\omega_j$ are the components of the angular
acceleration and angular velocity of the aggregate and the quantities $I_{ii}$
are the aggregate principal moments of inertia. These equations are solved
numerically at a given time step $t_n$ with the Boost Odeint C++ library using a
fifth-order Dormand-Prince Runge-Kutta method (with embedded fourth-order error
estimate) \cite{ahnertOdeintSolvingOrdinary2011}, yielding the instantaneous
angular velocity vector $\vec{\omega}(t_n)$. The time-step must ensure that the
rotation matrix $\overleftrightarrow{R}$ remains orthogonal and has determinant
equal to 1, guaranteeing proper physical rotations
\cite{goldsteinClassicalMechanics2008}. At each time step, the angular velocity
is recomputed from the current torques, and is then used to modify the position
of the aggregate by updating the positions of the principal axes using $\left(d
    \hat{R}_j / dt \right)_{\text{FS}} = \vec{\omega}^{\text{FS}} \times \hat{R}_j$.

\subsection{Interaction energy}
\label{subsec:Interaction energy}
Both conservative (sheath electric force, ion orbital force) and
non-conservative forces (ion collisional force, friction, Brownian force) act on
the aggregate. Consequently, the variation of the mechanical energy satisfies

\begin{equation}
    \Delta U + \Delta K =   W_{i,\text{coll}} + W_f + W_B.
\end{equation}

Here the quantities $\Delta U$ and $\Delta K$ denote the change in potential and
kinetic energies, while $W_{i,\text{coll}}$, $W_f$, and $W_B$ are the work done
on the aggregate by $\vec{F}^m_{i,\text{coll}}$, $\vec{F}^m_f$ and
$\vec{F}^m_B$.

As the aggregate rotates, the potential energy $U$ decreases toward a minimum
value, increasing the kinetic energy $K$. This kinetic energy is subsequently
dissipated by friction. Both the ion-collisional and Brownian work introduce
fluctuations in $K$. A rotational equilibrium is reached when the interaction
energy is minimized and the friction dissipates the excess kinetic energy.

Since the FS is centered at the aggregate center of mass, the gravitational
potential energy is zero; therefore the total potential energy calculated in the
FS is given by \cite{griffithsIntroductionElectrodynamics2013,
    jacksonClassicalElectrodynamics2009}

\begin{equation}
    \begin{aligned}
         & U = \sum_{m} Q_m \phi(\vec{r}_m) ,                                     \\
         & \phi(\vec{r}_m) = \phi_{\text{sheath}}(\vec{r}_m) + \phi_i(\vec{r}_m).
    \end{aligned}
\end{equation}
Here,  $\phi_{\text{sheath}}(\vec{r}_m)$ and $\phi_{i}(\vec{r}_m)$ are the
sheath electric potential and the ion electric potential evaluated at the
position $\vec{r}_m$ of monomer $m$.

Using $\vec{E}_{\text{sheath}} = -\vec{\nabla}\phi_{\text{sheath}}$, and the
electric dipole moment vector of the aggregate in the FS, Eq. (\ref{eq:dipole
    moment}), one obtains:

\begin{subequations}
    \begin{align}
         & U = U_p + U_i , \\
         & U =
        \underbrace{-\vec{p} \cdot \vec{E}_{\text{sheath}}}_{\text{dipole interaction}} +
        \underbrace{\textstyle\sum_{m} Q_m \phi_i(\vec{r}_m)}_{\text{ion interaction}}.
        \label{eq:real interaction energy}
    \end{align}
\end{subequations}

The ion-interaction contribution in Eq.~(\ref{eq:real interaction energy})
contains all the physical information inside the sum on $m$, but is unsuitable
for describing the system concisely. To make the contribution of different terms
explicit, we expand the ion potential $\phi_i(\vec{r}_m)$ around the aggregate
center of charge using a multipole expansion truncated at the second term:

\begin{equation}
    \begin{aligned}
         & \phi_i(\vec{r}_m) = \phi_{i}(\vec{r}_{\text{CQ}}) - \vec{r}_m\cdot \vec{E}_{i}(\vec{r}_{\text{CQ}}),                                                                                                               \\
         & \phi_{i}(\vec{r}_{\text{CQ}}) = \frac{1}{4\pi\epsilon_0} q_i \sum^{N_i}_{i=1} \exp{ \left (-|\vec{r}_{\text{CQ}} - \vec{r}_i| / \lambda_{D_e} \right )}/ |\vec{r}_{\text{CQ}} - \vec{r}_i|,                        \\
         & \begin{aligned}
               \vec{E}_{i}(\vec{r}_{\text{CQ}}) = & \frac{1}{4\pi\epsilon_0} q_i \sum^{N_i}_{i=1} \frac{1 + |\vec{r}_{\text{CQ}} - \vec{r}_i| / \lambda_{D_e}}{|\vec{r}_{\text{CQ}} - \vec{r}_i|^3} \\
                                                  & \exp{ \left (-|\vec{r}_{\text{CQ}} - \vec{r}_i| / \lambda_{D_e} \right )} (\vec{r}_{\text{CQ}} - \vec{r}_i)
           \end{aligned}
    \end{aligned}
\end{equation}

Substituting these expressions into Eq.~(\ref{eq:real interaction energy}), the
potential energy yields:

\begin{equation}
    U = -\vec{p}\,\cdot [\vec{E}_{\text{sheath}} + \vec{E}_i(\vec{r}_{\text{CQ}})] +
    Q \phi_i(\vec{r}_{\text{CQ}}).
    \label{eq:approximate interaction energy}
\end{equation}
In this expression, $\vec{E}_{i}(\vec{r}_{\text{CQ}})$, and $
    \phi_{i}(\vec{r}_{\text{CQ}})$, are the ion electric field and ion potential
evaluated at the aggregate center of charge ($\text{CQ}$) and $Q$ is the
aggregate total charge. The validity of this truncation is evaluated in
Appendix \ref{appx:validation in interaction energy}, where it is compared
with the direct calculation. Only 0.04\% of the data points exhibit a
relative error larger than 1\%.

This equation can be written in terms of the angles $\alpha_{p,i}$ between
$\vec{p}$ and the $\hat{x}$, $\hat{y}$, and $-\hat{z}$ directions in the FS:

\begin{equation}
    \begin{split}
        U = & -|\vec{p}\,| [E_{i,x} \cos{\alpha_{p,x}}+E_{i,y} \cos{\alpha_{p,y}} -                            \\
            & (\vec{E}_{\text{sheath}} + E_{i,\parallel}) \cos{\alpha_{p,z}}] + Q \phi_i(\vec{r}_{\text{CQ}}),
    \end{split}
    \label{eq:approximate interaction energy angles}
\end{equation}
Under laboratory conditions, the sheath electric field $\vec{E}_{\text{sheath}}$
is sufficiently strong to levitate the dust grains. Consequently, the
interaction energy is expected to be dominated by the sheath contribution. In
this regime, the dipole moment vector tends to align along the $-\hat{z}$
direction while the ion contribution acts as a perturbative correction.

\section{Results}

\label{sec:Results}

\subsection{Rotational dynamics for a representative aggregate}

To analyze the rotational dynamics, we focus on a representative irregular
aggregate made of melamine formaldehyde (mf) spheres, with physical properties
listed in Table~\ref{tab:agg_parameters}.  The aggregate's size is characterized
by its enclosing radius $R_{\text{enc}}$, defined as the maximum distance from
the center of mass to the outermost monomer. The plasma conditions correspond to
the sheath region in a GEC reference cell
(Table~\ref{tab:simulation_parameters}).

\begin{table}
    \centering
    \caption{Physical properties of the aggregate. $R_{\text{enc}}$ denotes the
        radius of the sphere enclosing the aggregate.}
    \begin{tabular}{|c|l|c|}
        \hline
        Symbol           & Parameter                   & Value
        \\
        \hline
        $N_m$            & Number of monomers          & $\mathrm{16}$
        \\
        $R_m$            & Monomer radius              & $\mathrm{4.46 \, \mu
                m}$
        \\
        $N_p$            & Patches per monomer         & $\mathrm{200}$
        \\
        $M_m$            & Monomer mass                & $\mathrm{5.63\times
                10^{-13}\,kg}$
        \\
        $M_{\text{agg}}$ & Aggregate mass              & $\mathrm{9.00\times
                10^{-12}\,kg}$
        \\
        $R_{\text{enc}}$ & Enclosing radius            & $\mathrm{36.5 \,\mu m}$
        \\
        $I_{xx}$         & Principal moment ($x$ axis) & $\mathrm{5.48\times
                10^{-22}\,kg\,m^2}$
        \\
        $I_{yy}$         & Principal moment ($y$ axis) & $\mathrm{3.88\times
                10^{-21}\,kg\,m^2}$
        \\
        $I_{zz}$         & Principal moment ($z$ axis) & $\mathrm{3.92\times
                10^{-21}\,kg\,m^2}$
        \\
        \hline
    \end{tabular}
    \label{tab:agg_parameters}
\end{table}

\begin{table}
    \centering
    \caption{Plasma parameters used in the simulations.}
    \begin{tabular}{|c|l|c|}
        \hline
        Symbol                & Parameter                        & Value
        \\
        \hline
        $P_n$                 & Neutral gas pressure             & $\mathrm{66\,
                Pa}$
        \\
        $n_0, n_e, n_i$       & Plasma, electron and ion density &
        $\mathrm{10^{14}\, m^{-3}}$
        \\
        $T_e$                 & Electron temperature             &
        $\mathrm{23400 \, K}$
        \\
        $T_i$                 & Ion temperature                  & $\mathrm{298
                \, K}$
        \\
        $T_n$                 & Neutral gas temperature          & $\mathrm{298
                \,K}$
        \\
        $E_{\text{sheath},z}$ & Sheath electric field            &
        $\mathrm{-6000\, V\,m^{-1}}$
        \\
        \hline
    \end{tabular}
    \label{tab:simulation_parameters}
\end{table}

We first describe the aggregate response, then investigate the cause of the
observed effects. First, the angular response and electric dipole evolution are
examined. Next, the torques responsible for the motion are studied. Finally, the
interaction energy is described. This sequence connects the  rotational
response, the mechanical causes,  and the energetic stabilization of the system.

\subsection{Rotational dynamics: angular velocity response and dipole alignment}

\begin{figure}
    \centering
    \includegraphics[width=\columnwidth]{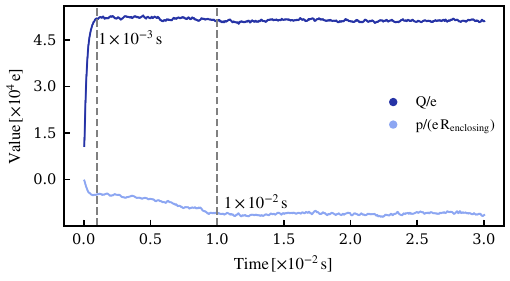}
    \caption{Temporal evolution of the aggregate's total charge ($Q/e$) and the
    normalized dipole moment magnitude ($|\vec{p}|/(e\,R_{\text{enc}})$). The
    beginning of the equilibrium phase for total charge and magnitude of dipole
    moment are denoted by dashed lines.}
    \label{fig:Q and p together}
\end{figure}

The simulation begins with an uncharged aggregate with random orientation and
ions flowing in the $-\hat{z}$ direction. As the aggregate charges, the
resulting torques cause the aggregate to rotate (see movie in supplementary
material). The dipole moment depends on both individual monomer charges and
their positions relative to the aggregate center of mass. As a result, its
evolution is affected by not only the charge redistribution but also by the
geometrical reorientation of the aggregate. Snapshots of the aggregate charge
distribution and the orientation of $\vec{p}$ at three representative time steps
during the rotational dynamics are shown in Fig.~\ref{fig:charge monomer
    orientation}. The aggregate's center of charge shifts as the aggregate rotates,
as a consequence of the gradual adjustment of the dipole magnitude.
Fig.~\ref{fig:Q and p together} shows the temporal evolution of $Q/e$ together
with the normalized magnitude of the dipole moment
$|\vec{p}|/(eR_{\text{enc}})$. While the total charge rapidly reaches a steady
value, the normalized dipole moment continues to evolve over a longer timescale.
This separation of timescales indicates that the alignment of $\vec{p}$ along
the $-\hat{z}$ direction is not determined by the total charge alone.

\begin{figure*}
    \centering
    \includegraphics[width=\textwidth]{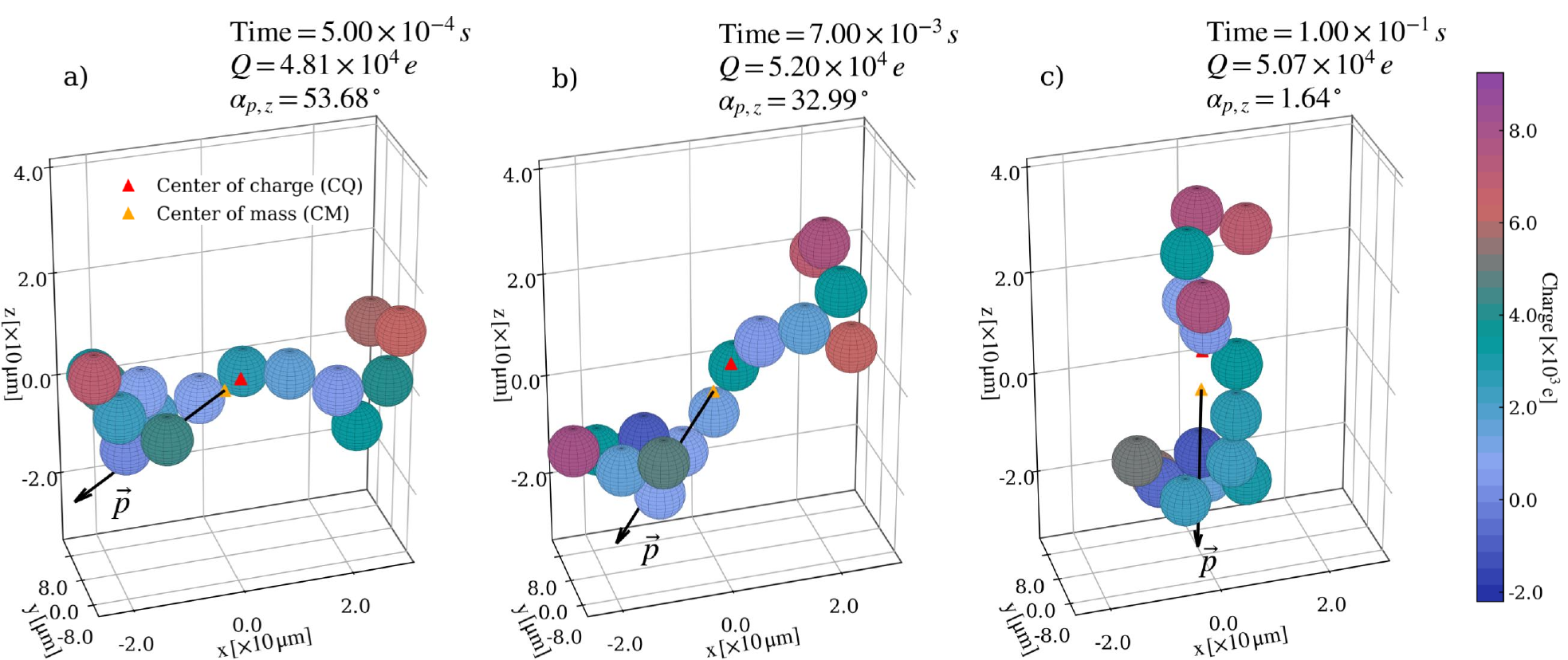}
    \caption{Evolution of monomer charge distributions at three representative
        times during rotational dynamics. Each panel shows the positions of the
        monomers (spheres), the center of mass (orange triangle), the center of
        charge (red triangle), and the electric dipole vector (arrow), where
        $\alpha_{p,z}$ denotes the angle between the dipole moment and the $-z$
        axis. A video is available in the supplementary material online.}
    \label{fig:charge monomer orientation}
\end{figure*}

\begin{figure}
    \centering
    \includegraphics[width=\columnwidth]{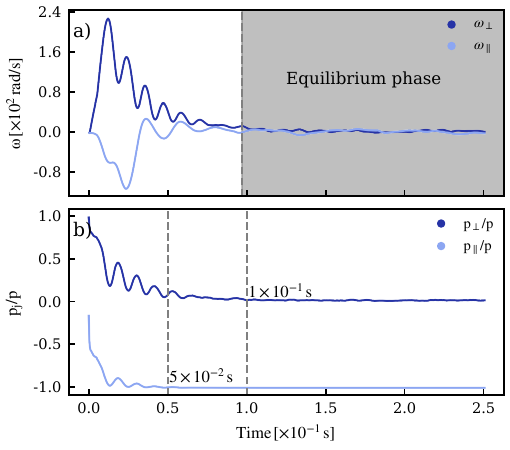}
    \caption{(a) Perpendicular ($\omega_{\bot }$) and parallel ($\omega_{\|}$)
        components of the aggregate's angular velocity during its rotational
        evolution. The equilibrium phase is represented by the shaded area and
        starts around the $t \sim 0.1\,s$. (b) Dynamical evolution of the
        normalized components of the aggregate's electric dipole moment
        ($p_{\parallel}/p$, $p_{\perp}/p$). Dashed lines indicate the onset of
        their equilibrium values.}
    \label{fig:angular velocity dipole}
\end{figure}

Fig.~\ref{fig:angular velocity dipole}a shows the parallel,
$\omega_{\|}=\omega_z$, and perpendicular, $\omega_{\bot }=\sqrt{\omega_x^2 +
        \omega_y^2}$, components of the angular velocity. After initial large
oscillations alter its orientation, both components ($\omega_{\bot }$ and
$\omega_{\|}$) decay towards zero. Although minor residual oscillations persist,
average rotation ceases, leaving the aggregate in a rotationally stable state.

\begin{figure}
    \centering
    \includegraphics[width=\columnwidth]{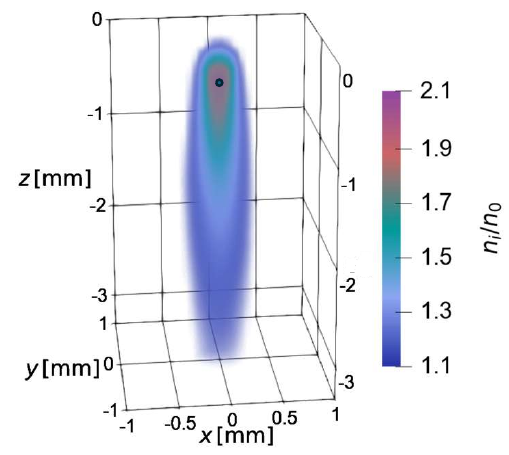}
    \caption{Averaged normalized ion density distribution in the equilibrium
    phase for a background plasma density of $n_0=10^{14}\,\mathrm{m}^{-3}$ and
    a sheath electric field of $E_{\text{sheath},z}=-6000\, \mathrm{V/m}$. The
    dot indicates the location of the aggregate's center of mass, scaled
    significantly larger than its enclosing radius ($R_{\text{enc}}$) for
    visualization purposes.}
    \label{fig:ion wake}
\end{figure}

\begin{figure}
    \centering
    \includegraphics[scale=1]{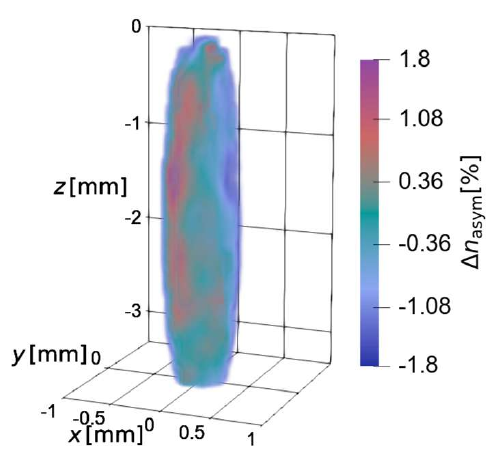}
    \caption{Relative asymmetry map of the averaged normalized ion density
        distribution in the equilibrium phase shown in Fig.~\ref{fig:ion wake}.
        This metric is computed as the relative difference between diametrically
        opposite points in the XY plane, defined as $\Delta n_{\text{asym}} [\%]
            = 100 \times (n_i - n_{i,\text{flip}})/n_i$.}
    \label{fig:ion wake differences}
\end{figure}

The ion environment around the aggregate at equilibrium is shown in
Fig.~\ref{fig:ion wake}. A region of enhanced ion density is observed downstream
of the aggregate, corresponding to the ion wake generated by the ion flow. The
extended wake structure establishes a highly asymmetric charge distribution
around the aggregate, but less obvious is the asymmetry about the $z-$axis (Fig.
\ref{fig:ion wake differences}).

\subsection{Torque balance and ionic field structure in the equilibrium phase}
\label{subsec: total aligning torque}

\begin{figure}
    \centering
    \includegraphics[width=\columnwidth]{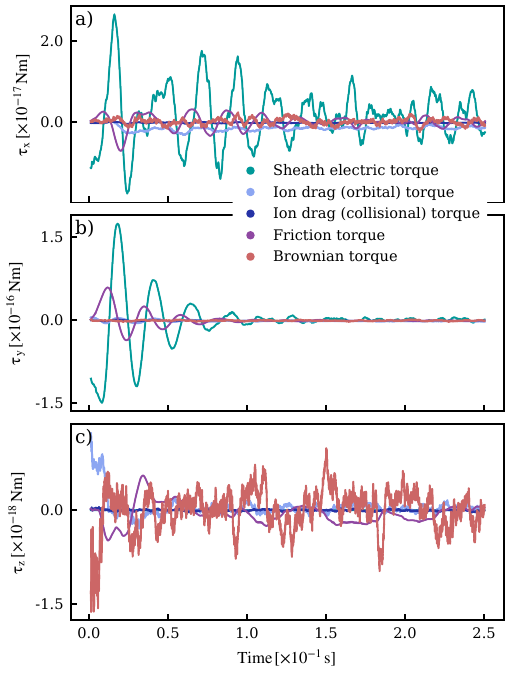}
    \caption{Components of $\vec{\tau}_e$, $\vec{\tau}_{i,\text{orb}}$,
    $\vec{\tau}_{i,\text{coll}}$, $\vec{\tau}_f$, and $\vec{\tau}_B$ torques
    acting on the aggregate during its rotational evolution. Panels show: (a)
    $x-$component, (b) $y-$component, and (c) $z-$component of the torque
    vectors. The simulation is performed with
    $E_{\text{sheath},z}=-6000\,\mathrm{V/m}$.}
    \label{fig:torques together}
\end{figure}

We now analyze the torques governing the aggregate rotation.
Fig.~\ref{fig:torques together} shows the temporal evolution of the $x-$, $y-$
and $z-$components of the torques due to the sheath electric field
($\vec{\tau}_e$), ion orbital-drag ($\vec{\tau}_{i,\text{orb}}$), ion
collisional-drag ($\vec{\tau}_{i,\text{coll}}$), neutral gas friction
($\vec{\tau}_f$) and Brownian ($\vec{\tau}_B$) forces acting on the aggregate.
The torques can be separated into two categories: (i) those that drive
reorientation ($\vec{\tau}_e$, $\vec{\tau}_{i, \text{orb}}$,
$\vec{\tau}_{i,\text{coll}}$) and (ii) those that modulate it ($\vec{\tau}_f$,
$\vec{\tau}_B$).

In the first $\sim 0.05\,s$, the $x-$ and $y-$components of $\vec{\tau}_e$ are
considerably larger than any other reorienting torque, confirming that
$\vec{\tau}_e$ is the dominant driver of the early rotational motion. Because
the sheath electric field $\vec{E}_{\text{sheath}}$ is applied along $-\hat{z}$,
the $z$-component of $\vec{\tau}_{e}$ is identically zero. Initially,
$\vec{\tau}_e$ and $\vec{\tau}_{i,\text{orb}}$ dominate the dynamics. As the
system evolves, $\vec{\tau}_{i, \text{orb}}$ becomes increasingly relevant,
while the $\vec{\tau}_e$ components are gradually damped as the aggregate
electric dipole moment vector $\vec{p}$ aligns with the direction of
$\vec{E}_{\text{sheath}}$.

Although the ion collisional-drag torque $\vec{\tau}_{i,\text{coll}}$ may
contribute to geometrically align the aggregate in the ion flow direction, its
magnitude is at least one order of magnitude smaller than contributions from
$\vec{\tau}_e$ and $\vec{\tau}_{i,\text{orb}}$. The friction torque
$\vec{\tau}_f$, proportional to the angular velocity $\vec{\omega}$, acts as a
dissipative mechanism that determines the time for the aggregate to reach
equilibrium.

\begin{figure}
    \centering
    \includegraphics[width=\columnwidth]{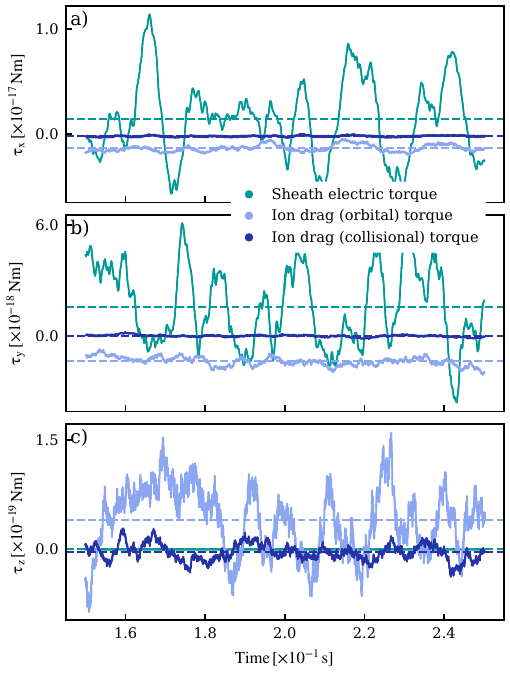}
    \caption{Components of $\vec{\tau}_e$, $\vec{\tau}_{i, \text{orb}}$, and
    $\vec{\tau}_{i, \text{coll}}$ acting on the aggregate during the equilibrium
    phase with a sheath electric field
    $E_{\text{sheath},z}=-6000\,\mathrm{V/m}$. Time-averaged torque components
    are represented by the dashed lines. Panels show: (a) $x-$component, (b)
    $y-$component, and (c) $z-$component of the torque vectors.}
    \label{fig:torques together equilibrium}
\end{figure}

After dynamic equilibrium is reached, marked by small perpendicular angular
velocity oscillations around zero (Fig.~\ref{fig:angular velocity dipole}a), the
mean aligning torques ($\vec{\tau}_e$, $\vec{\tau}_{i, \text{orb}}$, and
$\vec{\tau}_{i, \text{coll}}$) can be calculated. In this stationary regime
(Fig.~\ref{fig:torques together equilibrium}), the $x-$ and $y-$components of
$\vec{\tau}_e$ and $\vec{\tau}_{i, \text{orb}}$ dominate. Their opposite signs
reveal a sustained competition between these torques, establishing a dynamic
balance that maintains the aggregate's stable orientation.

The localized high ion density in the downstream wake (Fig.~\ref{fig:ion wake})
dominates the $\vec{\tau}_{i, \text{orb}}$ torque due to its proximity and
concentration. Since this wake extends predominantly along
$\vec{E}_{\text{sheath}}$, the resulting ion-generated electric field at the
aggregate is strongest in the $\hat{z}-$direction. Consequently, $\vec{\tau}_{i,
    \text{orb}}$ preferentially orients the aggregate along this axis.

In the following, we determine whether the ion influence on the aggregate
Eq.~(\ref{eq:approximate interaction energy}) can be described solely as a
dipole-ion electric field interaction, which would simplify the alignment
interpretation considerably. This assumption is only valid if the temporal
evolution of the dipole term ($-\vec{p}\cdot \vec{E}_i(\vec{r}_{\text{CQ}})$) is
much larger than the monopole term ($Q\phi_i(\vec{r}_{\text{CQ}})$) in the
equilibrium phase. To determine which term governs the rotational dynamics, we
compute their coefficients of variation $\mathrm{CV[\%]}=100
    \times(\sigma/|\mu|)$, where $\sigma$, $\mu$ are their standard deviations and
mean values. This magnitude quantifies how strongly the data fluctuate around
their mean value. A value of $\mathrm{CV}$ closer to zero corresponds to weaker
fluctuations in the data. The monopole and dipole term coefficients of variation
presented values of $0.760\%$ and $43.3\%$, respectively. As the monopole term
shows small variations during the equilibrium phase compared to the dipole term,
its contribution does not generate an appreciable torque on the aggregate, and
can be considered static in the stationary regime. Consequently, the total
aligning torque can be written as:

\begin{equation}
    \begin{aligned}
         & \vec{\tau}_{align} = \vec{\tau}_e + \vec{\tau}_{i, \text{orb}} \\
         & \vec{\tau}_{align} = \vec{p} \times \vec{E}_{\text{sheath}}  +
        \vec{p} \times \vec{E}_i(\vec{r}_{\text{CQ}})
    \end{aligned}
    \label{eq:aligning torques}
\end{equation}

The first term in the right-hand side represents the direct action of
$\vec{E}_{\text{sheath}}$ on the aggregate's dipole moment vector $\vec{p}$,
while the second term corresponds to the interaction of $\vec{p}$ with the ionic
electric field at the aggregate's center of charge
$\vec{E}_i(\vec{r}_{\text{CQ}})$.

\begin{figure}
    \centering
    \includegraphics[width=\columnwidth]{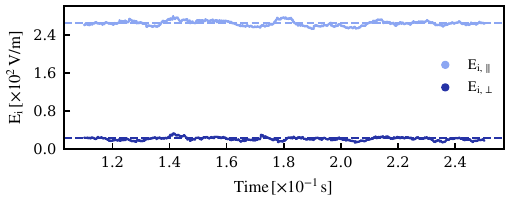}
    \caption{Components of $\vec{E}_i(\vec{r}_{\text{CQ}})$ acting on the
    aggregate during the equilibrium phase
    ($E_{\text{sheath},z}=-6000\,\mathrm{V/m}$). The dashed lines represent the
    time-averaged value of each component,  $\langle E_{i,\parallel}\rangle$ =
    264 V/m and $\langle E_{i,\perp}\rangle$ = 22 V/m.}
    \label{fig:ion electric field in time}
\end{figure}

Fig.~\ref{fig:ion electric field in time} shows the components of the ion
electric field $\vec{E}_i(\vec{r}_{\text{CQ}})$ evaluated at the center of
charge when the aggregate is in the equilibrium orientation. The components
remain approximately constant. $E_{i,\parallel}$ is the dominant component and
directed opposite to $\vec{E}_{\text{sheath}}$. The transverse component is
small but non-zero, an order of magnitude smaller than $E_{i,\parallel}$.  Its
influence becomes more pronounced when the dipole is perfectly aligned along the
$-\hat{z}$ direction. In this configuration, the torque associated with the
$z-$component of the total electric field ($\vec{E}_{\text{total},z} =
    \vec{E}_{\text{sheath}} + \vec{{E}}_{i,\parallel}$) vanishes, as it is collinear
with the dipole moment vector $\vec{p}$. Therefore, the smaller $E_{i,\perp}$
component can momentarily displace the aggregate from the alignment. As a result
of this perturbation, the $\vec{\tau}_e$, modulated by $\vec{E}_{i,\parallel}$,
which points opposite to $\vec{E}_{\text{sheath}}$, acts to restore the dipole
orientation, while friction dissipates the accumulated kinetic energy. Other
perturbations, such as the Brownian kicks and the ion collisions with the
aggregate surface also contribute to this process.

\begin{figure}
    \centering
    \includegraphics[width=\columnwidth]{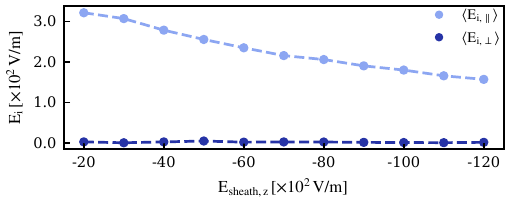}
    \caption{Time-averaged $x-$, $y-$ and $z-$components of
    $\vec{E}_i(\vec{r}_{\text{CQ}})$ for a spherical dust grain under different
    sheath electric field strengths. Note that  $\langle E_{i,\perp}\rangle$ is
    essentially zero.}
    \label{fig:spherical dust mean ion electric field}
\end{figure}

For comparison, Fig.~\ref{fig:spherical dust mean ion electric field} shows the
temporally averaged ion electric field components for the ion wake downstream of
a spherical grain under a range of sheath electric fields. The spherical grain
was chosen with a radius of $R_{\sigma}=14.42~\mu\mathrm{m}$, corresponding to
the equivalent radius of the aggregate
\cite{matthewsDiscreteStochasticCharging2018}. For a spherical grain, $\langle
    E_{i,\perp} \rangle$ is practically zero, and only the positive $\langle
    E_{i,\parallel} \rangle$ is significant. The cyclic competition between
misaligning and restoring torques can only emerge when a grain's geometry breaks
the symmetry of the wake, producing a non-zero $\langle E_{i,\perp} \rangle$,
i.e., for irregular aggregates. Note that for the same magnitude of the sheath
electric field ($E_{\mathrm{sheath}}=-6000~\mathrm{V/m}$), $\langle
    E_{i,\parallel} \rangle$ for the equivalent sphere is smaller than for the
aggregate ($234~\mathrm{V/m}$ compared with $264~\mathrm{V/m}$), consistent with
the lower average charge on the sphere, $4.35\times10^{4}e$, compared with
$5.1\times10^{4}e$ for the aggregate. The axial component of the ion wake
electric field decreases as the sheath electric field increases, due to a more
elongated ion wake.

\subsection{Interaction energy and potential well reconstruction}

After verifying the validity of the multipole expansion (Appendix
\ref{appx:validation in interaction energy}), the aggregate's interaction energy
with $\vec{E}_{\text{sheath}}$ and $\vec{E}_i(\vec{r}_{\text{CQ}})$ is examined.
In Eq.~(\ref{eq:approximate interaction energy angles}), $U$ depends on the
angles between $\vec{p}$ and the Cartesian directions. The system evolves to
minimize the interaction energy, but a stable configuration can only be reached
in the presence of a dissipative mechanism. Fig.~\ref{fig:interaction energy
    with alpha p} shows the time-dependent interaction energy $U$ as a function of
the angle of $\vec{p}$ and $\vec{E}_{\text{sheath}}$ (Eq.~(\ref{eq:approximate
    interaction energy})). As time progresses, $U$ approaches a minimum value for
$\alpha_{p,z}\approx 0^\circ$ indicating alignment with the sheath electric
field, whereas $\alpha_{p,x}$ and $\alpha_{p,y} \approx 90^\circ$ with average
deviation of $\sim 0.9^\circ$ due to the transverse components of
$\vec{E}_i(\vec{r}_{\text{CQ}})$.

\begin{figure}
    \centering
    \includegraphics[width=\columnwidth]{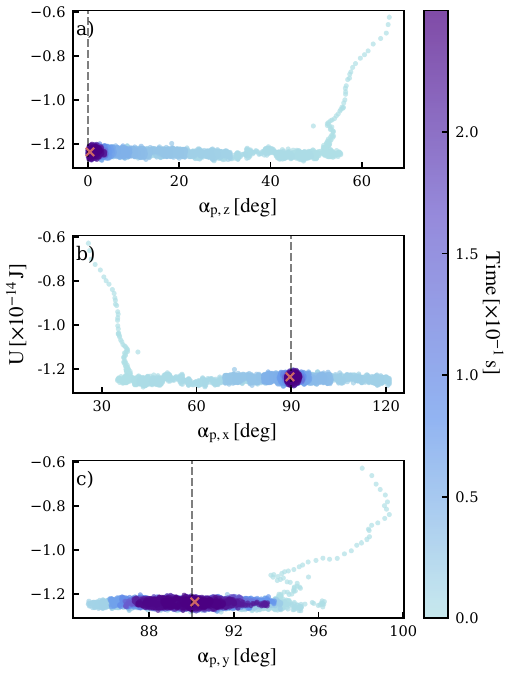}
    \caption{Time-dependent interaction energy for the electric dipole moment
    orientation angles with respect to $\vec{E}_{\text{sheath}}$. Panels show
    the interaction energy as a function of (a) $\alpha_{p,z}$, (b)
    $\alpha_{p,x}$, and (c) $\alpha_{p,y}$. Black dashed lines mark angles of
    $0^\circ$ in (a) and $90^\circ$ in (b) and (c). The orange cross indicates
    the mean interaction energy in the equilibrium phase $\langle U
        \rangle=-1.23 \times 10^{-14}\,\mathrm{J}.$}
    \label{fig:interaction energy with alpha p}
\end{figure}

To assess whether the rotational equilibrium state corresponds to true
mechanical stability rather than a transient alignment, we examine the effective
orientational potential. In this configuration, the aggregate lies in a
potential well where the angle between the $\vec{p}$ and
$\vec{E}_{\text{sheath}}$ exhibits small oscillations around its mean value.
Because the simulation employs a super-ion model, the interaction energy data
can be quite noisy. To reduce this effect, the data in the equilibrium window
were grouped into angular bins. The probability of an aggregate orientation
angle to be in a bin centered at $\alpha_{p,z}$ is:

\begin{equation}
    P(\alpha_{p,z}) = \frac{N(\alpha_{p,z})}{\sum_{\text{bins}} N(\alpha_{p,z})} ,
\end{equation}
where $N(\alpha_p{,z})$ is the number of counts per bin and $\sum_{\text{bins}}
    N(\alpha_{p,z})$ is the total number of counts in the equilibrium phase.

To reconstruct the underlying potential well, the canonical ensemble formulation
is applied \cite{greinerThermodynamicsStatisticalMechanics1995,
    pathriaStatisticalMechanics2011}. For this case, the aggregate is treated as the
ensemble while the surrounding plasma acts as the heat bath. The probability of
a microstate is given by:

\begin{equation}
    P(\alpha_{p,z}) = \frac{\exp(- \frac{H(\alpha_{p,z}) }{\beta})}{Z} .
    \label{canonical ensemble probability}
\end{equation}

Here, $H(\alpha_{p,z}) = K(\alpha_{p,z})  + U(\alpha_{p,z})$. The quantities
$H(\alpha_{p,z})$, $K(\alpha_{p,z})$ and $U(\alpha_{p,z})$ denote the
bin-averaged Hamiltonian, kinetic energy and interaction energy over the angular
interval centered at $\alpha_{p,z}$, respectively. The parameter
$\beta=\frac{3}{2 \langle K \rangle}$, where $\langle K \rangle$ is the
time-averaged kinetic energy in the equilibrium phase and $Z =
    \sum_{\alpha_{p,z}} \exp(- \frac{H(\alpha_{p,z}) }{\beta})$ is the canonical
ensemble partition function. Taking the natural logarithm of Eq.~(\ref{canonical
    ensemble probability}), the normalized total interaction energy $U \beta$ can be
obtained as:

\begin{equation}
    U(\alpha_{p,z})\beta = -\beta(\frac{\ln(P(\alpha_{p,z}))}{\beta} + K(\alpha_{p,z}))
    \label{eq:interaction energy canonical ensemble}
\end{equation}
where the constant term $\beta \ln(Z)$ was absorbed into $U(\alpha_{p,z})$.

\begin{figure}
    \centering
    \includegraphics[width=\columnwidth]{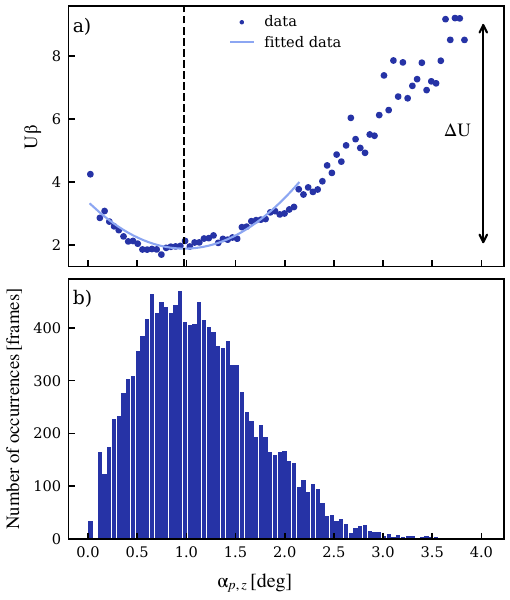}
    \caption{Potential well and angular orientation distribution for an
        aggregate in the sheath electric field during the rotational equilibrium
        phase. Panel (a) shows the normalized interaction energy (points) with a
        dashed line indicating the position where the potential energy is a
        minimum $U_{\text{min}}$. A parabolic fit around the minimum is used to
        determine the restoring torque coefficient $k$, while $\Delta U$
        represents the well depth. Panel (b) shows the angular occurrence
        frequency.}
    \label{fig:potential well}
\end{figure}

The total interaction energy $U$ as a function of the orientation angle
$\alpha_\text{p, z}$ is shown in Fig.~\ref{fig:potential well}a.
Fig.~\ref{fig:potential well}b displays the number of occurrences in each
angular bin. Given the aggregate is in an equilibrium configuration, it should
remain longer in orientations that minimize $U$. Therefore, the angles
corresponding to minimum energy appear more frequently in the angular
distribution.

The stability of the potential well is quantified by two constants, the spring
constant $\kappa$, which measures the potential curvature around the equilibrium
angle, and the well depth $\Delta U$, which determines the energy barrier that
the aggregate must overcome to escape from the well. Physically, $\kappa$
quantifies the restoring torque and therefore the rotational stiffness of the
aggregate. $\Delta U$, on the other hand, defines the energy threshold for
confinement: the aggregate remains bound as long as its total energy does not
exceed this barrier. The spring constant $\kappa$ is computed from a parabolic
fit around the energy minimum, $U=U_0 + \frac{1}{2}\kappa(\alpha_{p,z} -
    \alpha_{p,z_0})^2$, and the energy well depth is obtained as $\Delta U =
    U_{\text{max}} - U_{\text{min}}$.

In the next subsection \ref{subsec:spring constant and energy well depth}, a
detailed analysis is performed for multiple aggregates under three different
sheath electric field intensities. The corresponding values of $\kappa$ and
$\Delta U$ are presented in that section.

\subsection{Average dipole moment in the equilibrium phase}

\begin{figure*}
    \centering
    \includegraphics[width=\textwidth]{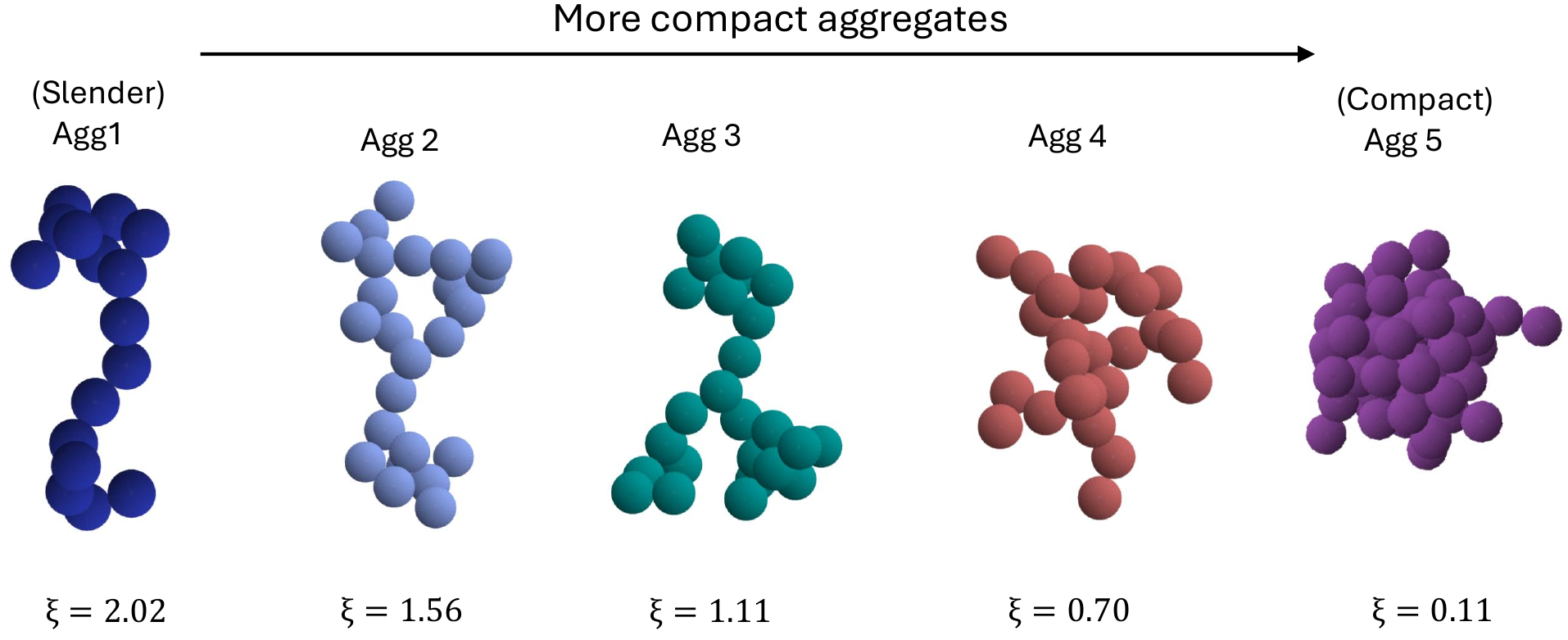}
    \caption{Irregular aggregates ranging from slender to nearly spherical
        shapes. The elongation factor, calculated according to
        Eq.~\ref{eq:elongation factor}, is used to distinguish the shape of the
        aggregates but is not predictive to their charge or dynamics.}
    \label{fig:aggs}
\end{figure*}

\begin{table}
    \centering
    \caption{Physical properties for the five aggregates studied, where
        $R_{\text{enc}}$ is the enclosing radius and $I_{ii}$ represents the
        principal moments of inertia. The monomer radius and mass are $R_m =
            4.46 \times 10^{-6}\,\mathrm{m}$ and $M_m = 5.63 \times
            10^{-13}\,\mathrm{kg}$, respectively. Aggregate 1 (Agg 1) corresponds to
        the reference configuration analyzed in the previous sections.}

    \begin{tabular}{|c|c|c|c|c|c|}
        \hline
        Symbol                                           & Agg1            &
        Agg2                                             & Agg3            &
        Agg4                                             & Agg5              \\
        \hline
        $\xi$                                            & $\mathrm{2.02}$ &
        $\mathrm{1.56}$                                  & $\mathrm{1.11}$ &
        $\mathrm{0.70}$                                  & $\mathrm{0.11}$   \\
        $N_m$                                            & $\mathrm{16}$   &
        $\mathrm{24}$                                    & $\mathrm{25}$   &
        $\mathrm{25}$                                    & $\mathrm{64}$     \\
        $M_{\text{agg}}\,[\times \mathrm{10^{-11}\,kg}]$ & $\mathrm{0.9 }$ &
        $\mathrm{1.31}$                                  & $\mathrm{1.40}$ &
        $\mathrm{1.40}$                                  & $\mathrm{3.60}$   \\
        $R_{\text{enc}}\,[\mathrm{\mu m}]$               & $\mathrm{36.5}$ &
        $\mathrm{39.5}$                                  & $\mathrm{38.4}$ &
        $\mathrm{34.7}$                                  & $\mathrm{35}$     \\
        $I_{xx}\,[\times \mathrm{10^{-21}\,kg\,m^2}]$    & $\mathrm{0.54}$ &
        $\mathrm{1.40}$                                  & $\mathrm{2.07}$ &
        $\mathrm{2.13}$                                  & $\mathrm{7.27}$   \\
        $I_{yy}\,[\times \mathrm{10^{-21}\,kg\,m^2}]$    & $\mathrm{3.88}$ &
        $\mathrm{6.14}$                                  & $\mathrm{5.34}$ &
        $\mathrm{2.90}$                                  & $\mathrm{7.36}$   \\
        $I_{zz}\,[\times \mathrm{10^{-21}\,kg\,m^2}]$    & $\mathrm{3.92}$ &
        $\mathrm{6.83}$                                  & $\mathrm{6.20}$ &
        $\mathrm{3.84}$                                  & $\mathrm{7.41}$   \\
        \hline
    \end{tabular}
    \label{tab:aggs_parameters}
\end{table}

To extend the results obtained for a single aggregate to a broader set of
conditions and aggregate shapes, we examined five different aggregates
(including the one previously analyzed)  with varying elongation parameters (see
Fig.~\ref{fig:aggs}). Physical characteristics of the aggregates are listed in
Table \ref{tab:aggs_parameters}. These aggregates were studied individually at
three different sheath electric field strengths: $-6000$, $-9000$, and
$-12000~\mathrm{V/m}$.

As shown in Fig.~\ref{fig:angular velocity dipole}b, the alignment of the dipole
moment vector $\vec{p}$ along the direction of $\vec{E}_{\text{sheath}}$
corresponds to the system's minimum energy configuration.
Table~\ref{tab:mean_inclination_for_aggs} shows the average angular deviation of
the dipole moment from the $-\hat{z}$ direction after dynamic equilibrium is
reached. For each case, $\vec{p}$ is well-aligned with a deviation less than
three degrees, indicating that the system is in a stable configuration. The
dipole inclination does not show a clear trend with increasing electric field
intensity. As the electric field increases, the collected charge and the dipole
moment components also increase (see Table~\ref{tab:aggs_Q_p}), so the final
orientation can be slightly different (less than 0.6$^{\circ}$).  Note that the
dipole moment is not always aligned with the principal axis associated with the
smallest moment of inertia.  In general, the alignment is better for long
slender aggregates (e.g. values of $\langle\alpha_{p,x'}\rangle$ for Agg1 and
Agg2 in Table \ref{tab:mean_inclination_for_aggs}). However, note the large
deviation in $\langle\alpha_{p,x'}\rangle$ for Agg3, with $\xi = 1.11$, and
relatively good alignment for the more compact Agg4, with $\xi = 0.70$.

\begin{table}
    \centering
    \caption{Time-averaged normalized charge and dipole moment for the five
    aggregates. All values are reported in units of elementary charge ($e$) and
    scaled by a factor of $10^4$. $Q_0=\langle Q/e \rangle$ and $p_0=\langle
        p/(|e|\,R_{\text{enc}})\rangle$.}
    \label{tab:aggs_Q_p}
    \begin{tabular}{|c|cc|cc|cc|}
        \hline
        \multirow{2}{*}{Aggs}                     &
        \multicolumn{2}{c|}{$-6000~\mathrm{V/m}$} &
        \multicolumn{2}{c|}{$-9000~\mathrm{V/m}$} &
        \multicolumn{2}{c|}{$-12000~\mathrm{V/m}$}                            \\
        \cline{2-7}
                                                  & $Q_0$  & $p_0$  & $Q_0$ &
        $p_0$                                     & $Q_0$  & $p_0$            \\
        \hline
        Agg1                                      & $5.1$  & $1.1$  & $5.7$ &
        $1.2$                                     & $6.1$  & $1.3$            \\
        Agg2                                      & $6.2$  & $1.2$  & $6.9$ &
        $1.3$                                     & $7.4$  & $1.4$            \\
        Agg3                                      & $6.3 $ & $0.82$ & $7.1$ &
        $0.94$                                    & $7.7$  & $1.0$            \\
        Agg4                                      & $6.1$  & $0.89$ & $6.8$ &
        $0.98$                                    & $7.4$  & $1.0$            \\
        Agg5                                      & $7.7$  & $1.5$  & $8.5$ &
        $1.6$                                     & $9.1$  & $1.6$            \\
        \hline
    \end{tabular}
\end{table}

\begin{table}
    \centering
    \caption{Mean equilibrium dipole inclination  in degrees with respect to the
        $-\hat{z}$ axis, $\langle \alpha_{p,z} \rangle$, and the
        $\hat{x}_\text{BS}$ axis, $\langle \alpha_{p,x'} \rangle$.}
    \begin{tabular}{|c|cc|cc|cc|}
        \hline
        \multirow{2}{*}{Aggs}                     &
        \multicolumn{2}{c|}{$-6000~\mathrm{V/m}$} &
        \multicolumn{2}{c|}{$-9000~\mathrm{V/m}$} &
        \multicolumn{2}{c|}{$-12000~\mathrm{V/m}$}
        \\
        \cline{2-7}
                                                  & $\langle \alpha_{p,z}
        \rangle$                                  & $\langle \alpha_{p,x'}
        \rangle$                                  & $\langle \alpha_{p,z} \rangle$ & $\langle
        \alpha_{p,x'} \rangle$                    & $\langle \alpha_{p,z} \rangle$ &
        $\langle \alpha_{p,x'} \rangle$
        \\
        \hline
        Agg1                                      & $1.14$
                                                  & $3.27$                         & $1.15$   & $3.17$ & $1.34$ & $3.17$ \\
        Agg2                                      & $1.36$
                                                  & $5.08$                         & $1.41$   & $5.18$ & $1.30$ & $5.76$ \\
        Agg3                                      & $2.96$
                                                  & $41.2$                         & $2.39$   & $47.0$ & $2.30$ & $50.6$ \\
        Agg4                                      & $1.97$
                                                  & $4.51$                         & $2.16$   & $4.63$ & $2.31$ & $6.84$ \\
        Agg5                                      & $1.61$
                                                  & $48.5$                         & $1.78$   & $48.0$ & $2.13$ & $47.9$ \\
        \hline
    \end{tabular}
    \label{tab:mean_inclination_for_aggs}
\end{table}

\subsection{Mean ionic electric field components}

\begin{figure}
    \centering
    \includegraphics[width=\columnwidth]{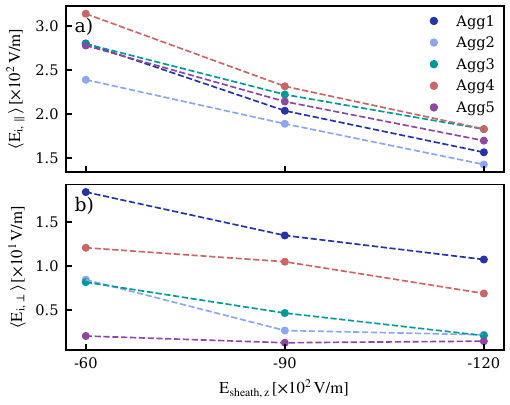}
    \caption{Time-averaged ion electric field components for five different
        aggregates at different strengths of the sheath electric field. The
        parallel and perpendicular components are shown in panels (a), (b),
        respectively. Dashed lines serve to guide the eye.}
    \label{fig:mean ion electric field}
\end{figure}

We assess how the ion electric field contributes to the aligning torques
(Eq.~(\ref{eq:aligning torques})). For all five aggregates, and across the three
sheath electric field strengths, the time-averaged components of
$\vec{E}_i(\vec{r}_{\text{CQ}})$ were evaluated during the equilibrium phase.
Since the aligning torques depend on the term
$\vec{p}\times\vec{E}_i(\vec{r}_{\text{CQ}})$, it is essential to determine
whether the behavior identified for a single aggregate extends to a broader set
of shapes (see Fig.~\ref{fig:mean ion electric field}). In all cases, $\langle
    E_{i,\parallel} \rangle$ remains the dominant component, positive and opposing
to $\vec{E}_{\text{sheath}}$ direction. The transverse component $\langle
    E_{i,\perp} \rangle$ is $\sim10 \times$ smaller than the averaged axial
component $\langle E_{i,\parallel} \rangle$ but nonzero due to the aggregates'
asymmetry. Furthermore, as the sheath electric field increases, the magnitude of
$\langle E_{i,\perp} \rangle$ is observed to decrease; the stronger field leads
to a more elongated ion wake that suppresses transverse asymmetries in the local
electric field. No real trend with shape was observed, except that nearly
spherical aggregate has $\langle E_{i,\perp} \rangle \sim 0$. The consistent
pattern of a stronger axial component accompanied by weaker transverse
components generalizes the behavior observed for the single aggregate analysis.

\subsection{Configuration energy}
\label{subsec:spring constant and energy well depth}

\begin{figure}
    \centering
    \includegraphics[width=\columnwidth]{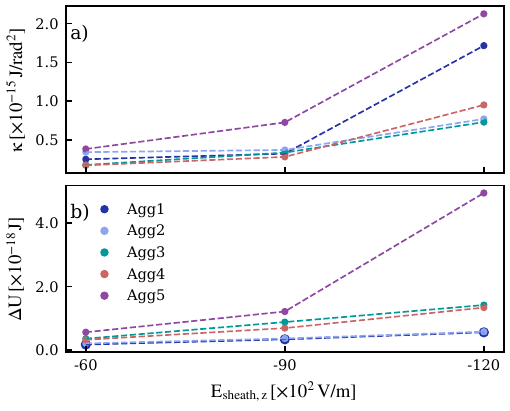}
    \caption{(a) Variation of the spring constant $k$ and (b) variation of the
        energy well depth $\Delta U$ as a function of the sheath electric field
        strength for the five aggregates analyzed.}
    \label{fig:spring constant and energy well depth}
\end{figure}

As introduced in the analysis of a single aggregate, the spring constant
$\kappa$ and the interaction energy well depth $\Delta U$ quantify complementary
aspects of the rotational equilibrium phase: $\kappa$ measures the local
stability of the potential well, whereas $\Delta U$ represents the potential
energy threshold required for the system to escape its local equilibrium state.
Fig.~\ref{fig:spring constant and energy well depth} presents these quantities
for all five aggregates at the three values of $E_{\text{sheath}}$ considered.
Both $\kappa$ and $\Delta U$ increase markedly with $E_{\text{sheath}}$ across
all cases, indicating that increasing $E_{\text{sheath}}$ enhances the
rotational alignment and reduces the angular oscillations about the equilibrium
position. Thus, the sheath electric field magnitude is the dominant factor
controlling the stability of the rotational equilibrium for a given aggregate
and therefore plays a central role in the aggregate dynamics. A systematic
assessment of shape-dependent differences would require a larger ensemble of
aggregates.

\section{Conclusions}
The rotational dynamics of irregular dust aggregates immersed in the sheath of a
GEC cell plasma chamber were analyzed using a self-consistent numerical
approach. The results show that aggregate orientation is mainly governed by
competition between the sheath electric ($\vec{\tau}_e$) and ion drag (orbital)
($\vec{\tau}_{i,\text{orb}}$) torques. These combined effects gradually orient
the dipole moment vector of the aggregate along the direction of
$\vec{E}_{\text{sheath}}$, establishing an equilibrium configuration. After
dynamic equilibrium is reached, the non-zero transverse components of ion
electric field, produced by the asymmetry of the aggregate, cause a slight
misalignment of the aggregate dipole moment and the sheath electric field. The
maximum misalignment is less than three degrees for the aggregates considered
(Table \ref{tab:mean_inclination_for_aggs}). A second-order multipole expansion
shows that the dipolar term dominates the ions' contribution to the aligning
torque, making the dipole-ion interaction a robust approximation across all
examined conditions.

At equilibrium, the configuration energy can be described by a potential energy
well whose depth and spring constant increase with the $E_{\text{sheath}}$. For
stronger fields, the systems present a better rotational alignment and are more
resilient to angular perturbations.

Altogether, these results indicate that the sheath electric field acting on an
aggregate's dipole moment is the main driver of rotation and the principal
stabilizing factor in the equilibrium phase. The axial component of the ion-wake
field produces an opposing aligning torque, while its transverse components act
as a destabilizing mechanism that leads to small oscillations about the
equilibrium position. However, for any given aggregate geometry, it remains
difficult to predict the dipole moment and its alignment with the principal
axes.

\section*{Acknowledgments}

The authors gratefully acknowledge support from the US Department of Energy,
Office of Science, Office of Fusion Energy Sciences under award number
DE-SC0024681 and the National Science Foundation grant PHY 2308743.

\appendix

\section{}
\label{appx:validation in interaction energy}

\begin{figure}[h]
    \centering
    \includegraphics[width=\columnwidth]{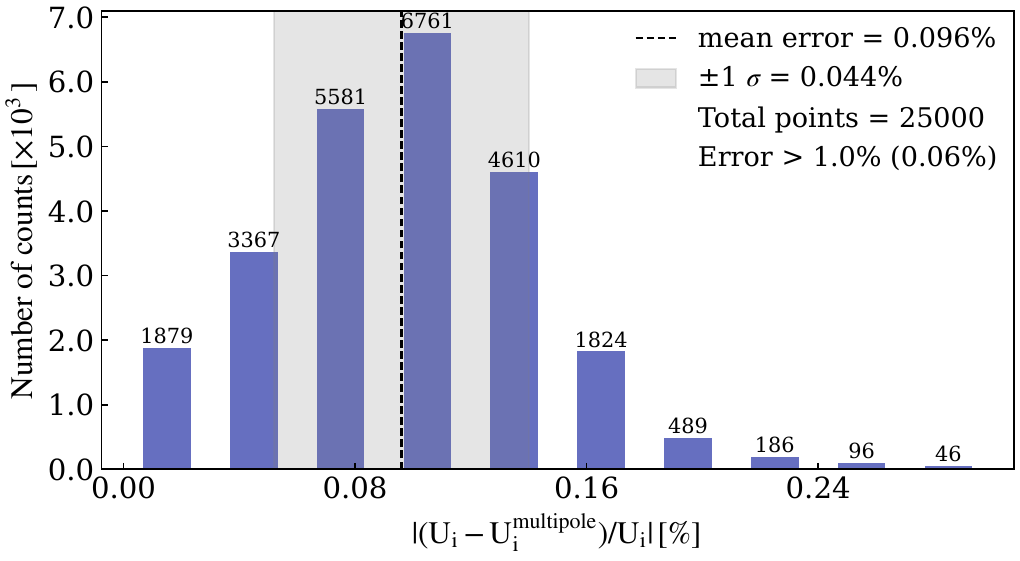}
    \caption{Relative error between the direct calculation of the ion
        interaction energy (Eq.~(\ref{eq:real interaction energy})) and its
        multipole expansion approximation (Eq.~(\ref{eq:approximate interaction
            energy})).}
    \label{fig:Error in U}
\end{figure}

It is necessary to verify whether the multipole expansion of the ion interaction
energy is suitable to describe our system. Fig.~\ref{fig:Error in U} shows the
relative difference between the terms $U_i$ calculated directly through
Eq.~(\ref{eq:real interaction energy}) and obtained with the multipole expansion
in Eq.~(\ref{eq:approximate interaction energy}). The histogram indicates that
more than 99\% (24 840) of the 25 000 time steps analyzed present errors below
1\%, while the outliers represent less than 0.04\% of the data.

This confirms the validity of the multipole expansion for the aggregate's total
interaction energy (Eq.~(\ref{eq:approximate interaction energy})), allowing the
angular evolution to be described in terms of the interaction between the dipole
moment vector and the total electric field at the aggregate's center of charge.

The relative error between the exact calculation of the interaction energy and
its multipole approximation was analyzed for the five different aggregates.
Table~\ref{tab:percent outliers for aggs} summarizes how frequently errors
exceed 1\% for the different aggregates and sheath electric field strengths. In
all cases, the percent of outliers is less than 0.5\%, confirming the multipole
expansion remains valid even for different aggregate shapes and sheath electric
field strengths.

\begin{table}[h]
    \centering
    \caption{Percentage of outliers with relative error greater than 1\% for
        five aggregates at three sheath electric field strengths.}
    \begin{tabular}{|c|c|c|c|}
        \hline
        Aggs & $-6000 \mathrm{V/m}$ & $-9000 \mathrm{V/m}$ & $-12000
        \mathrm{V/m}$                                                \\
        \hline
        Agg1 & 0.0560               & 0.0360               & 0.0360  \\
        Agg2 & 0.220                & 0.148                & 0.104   \\
        Agg3 & 0                    & 0                    & 0       \\
        Agg4 & 0.0720               & 0.0520               & 0.0480  \\
        Agg5 & 0.0160               & 0.0240               & 0.0120  \\

        \hline
    \end{tabular}
    \label{tab:percent outliers for aggs}
\end{table}

We calculate the Coefficient of Variation (CV) for the ion interaction monopole
term (Eq.~(\ref{eq:approximate interaction energy})) and the dipole term to
verify that Eq.~(\ref{eq:aligning torques}) is still applicable. As shown in
Table~\ref{tab:variation coefficient for aggs}, the CV for the monopole term
remains nearly constant ($\mathrm{CV} < 0.81\%$) across all cases, whereas the
dipole term exhibits significantly larger fluctuations, reinforcing that the
rotational dynamics are dominated by dipolar interactions.

\begin{table}
    \centering
    \caption{Coefficient of variation ($\mathrm{CV}$) for the monopole
    ($Q\phi_i(\vec{r}_{\text{CQ}})$) and dipole
    ($-\vec{p}\cdot\vec{E}_i(\vec{r}_{\text{CQ}})$) terms across the five
    aggregates shown in Fig.~\ref{fig:aggs} and the sheath electric field
    strengths.}
    \begin{tabular}{|c|cc|cc|cc|}
        \hline
        \multirow{2}{*}{Aggs}                     &
        \multicolumn{2}{c|}{$-6000~\mathrm{V/m}$} &
        \multicolumn{2}{c|}{$-9000~\mathrm{V/m}$} &
        \multicolumn{2}{c|}{$-12000~\mathrm{V/m}$}
        \\
        \cline{2-7}
                                                  & Monopole & Dipole   &
        Monopole                                  & Dipole   & Monopole & Dipole
        \\
        \hline
        Agg1                                      & 0.767    & 46.0     & 0.799
                                                  & 57.9     & 0.801    & 68.7
        \\
        Agg2                                      & 0.697    & 242      & 0.715
                                                  & 133      & 0.752    & 121
        \\
        Agg3                                      & 0.678    & 35.4     & 0.729
                                                  & 34.9     & 0.756    & 31.9
        \\
        Agg4                                      & 0.707    & 62.5     & 0.723
                                                  & 118      & 0.755    & 107
        \\
        Agg5                                      & 0.650    & 56.4     & 0.686
                                                  & 45.3     & 0.692    & 45.0
        \\
        \hline
    \end{tabular}
    \label{tab:variation coefficient for aggs}
\end{table}

\bibliographystyle{apsrev4-2}
\bibliography{Library_1}

\end{document}